\documentclass[amstex,epsf,amssymb,usenatbib]{mnras}
\usepackage{epsf,graphics,subfig}
\usepackage{amsmath,amssymb,natbib}
\usepackage{lmodern}
\usepackage{cite}

\usepackage{rotating,times,graphicx,latexsym,color,longtable,lscape} 

\def\aj{AJ}                   
\def\araa{ARA\&A}             
\def\apj{ApJ}                 
\def\apjl{ApJ}                
\def\apjs{ApJS}               
\def\apss{Ap\&SS}             
\def\aap{A\&A}                
\def\mnras{MNRAS}             
\def\pasj{PASJ}               

\newcommand{\degsy}{$^\circ$}

\title[X-ray cavities of radio galaxies]{Three-dimensional simulations of X-ray cavities inflated by radio galaxies.}
    
\author[M.D. Smith. \& J. Donohoe]
{ Michael D. Smith $^{1}$\thanks{E-mail: m.d.smith@kent.ac.uk} \&
{Justin Donohoe $^{1}$\thanks{E-mail: jd440@kent.ac.uk, }  }\\
$^{1}$Centre for Astrophysics \& Planetary Science, The University of Kent, Canterbury, Kent CT2 7NH, U.K. }                                                                                                                                                             
\date{Accepted .....
      Received ..... ;
      in original form .....}

\pagerange{\pageref{firstpage}--\pageref{lastpage}}
\pubyear{2019}

\begin{document}
                                                                                                                                                             
\maketitle
                                                                                                                                                             
\label{firstpage}
                                                                                                                                                             
\begin{abstract}
Vast cavities in the intergalactic medium are excavated by radio galaxies. The cavities appear as such in X-ray images because the external medium has been swept up, leaving a hot but low density bubble surrounding the radio lobes.  We explore here  the predicted thermal X-ray emission from a large set of   high-resolution three dimensional simulations of radio galaxies driven by supersonic jets.
We assume adiabatic non-relativistic hydrodynamics with injected straight and precessing jets of supersonic gas emitted from nozzles. 
Images of X-ray Bremsstrahlung emission  tend to generate oval cavities in the soft keV bands and leading arcuate structures in hard X-rays. However, the cavity shape is sensitive to the jet-ambient density contrast, varying from concave-shaped at $\eta = 0.1$ to convex for $\eta = 0.0001$ where $\eta$ is the jet/ambient density ratio. 
 We find  lateral ribs in the soft X-rays  in certain cases and propose this as an  explanation for  those detected in the vicinity of Cygnus\,A. In bi-lobed or X-shaped sources and in curved or deflected jets, the strongest X-ray emission is not associated with the hotspot but with the relic lobe or deflection location. This is because the hot high-pressure and dense high-compression regions do not coincide. Directed toward the observer, the cavity becomes a deep round  hole surrounded by circular ripples.  
With short radio-mode outbursts with a duty cycle of 10\% , the intracluster medium simmers with low Mach number shocks widely dissipating the jet energy in between active jet episodes. 
\end{abstract} 
  
\section{Introduction} 
\label{intro}

It is well established that powerful Active  Galactic Nuclei are capable of inflating bubbles in the intergalactic and intracluster media. This includes vast cavities created when gas is driven out by powerful radio galaxies. The cavities are observed as depressions in the surface brightness of X-ray emission which encompass the radio  lobes. Such cavities are found in  diverse cluster environments and surround a variety of radio galaxy types  \citep{1994MNRAS.270..173C,2011ApJ...732...95G,2012MNRAS.421..808P,2017MNRAS.466.2054V,2019ApJ...870...62P}; with evidence for low Mach number bow shocks being driven into, and capable of supporting, the  environment \citep{2006ApJ...644L...9W,2014ApJ...794..164S}. 
This phenomenon takes on extra significance as we probe the regulatory role it has on galaxy formation through feedback from Active  Galactic Nuclei into the cluster gas \citep{2012ARA&A..50..455F,2019A&A...622A..12H}. 

With the advent of new X-ray technology, we can now expect detailed measurements of the structure of these cavities at multiple frequencies. 
Our main purpose here is to predict how this X-ray structure would appear if the radio galaxy is driven by supersonic hydrodynamic jets. We aim to account for the Mach number, precession and orientation in order to provide a rich bank of simulated images. The physical properties and radio morphologies have already been published for this set of three-dimensional simulations in
Paper\,1 \citep{2016MNRAS.458..558D} and Paper\,2 \citep{2019MNRAS.490.1363S}. We thereby wish to  probe the implications for the intracluster medium. 
Specifically, these cavities may provide the link between the regulation of star formation in central massive galaxies and cluster cooling flows. In this respect, we already confirmed in Paper 1  that a high fraction of the jet energy is transferred and dissipated as thermal energy into the intracluster gas.  

An X-ray cavity was first associated with the radio source 3C\,84 in the Perseus elliptical galaxy NGC\,1275 by \citet{1993MNRAS.264L..25B}. The Perseus intracluster medium was found to be displaced by the radio lobes although the minimum lobe pressure falls below that of the thermal gas. The brightest regions are located just outside the radio lobes. In addition, the jets appear to change angle from the south-east towards the south with the lobes extending towards the west. While this may indicate jet deflection, it could also be directly related to jet precession. Note that the X-ray ROSAT image is from the soft range between 0.1 and 2.4\,keV.

Another X-ray cavity was then  associated with the  radio galaxy Cygnus\,A by \citet{1994MNRAS.270..173C}. Such X-ray cavities are likely to endure, remaining intact for over 100\,Myr, well after the high energy relativistic electrons have cooled within the lobes. Superimposed X-ray structure can also take the form of ripples, edges, bow shocks and cold fronts \citep[e.g.][]{2004ApJ...607..800B,2011ApJ...734L..28C,2016ApJS..227...31S,2015Ap&SS.359...61S}.   

Numerical simulations of jet-driven radio galaxies have shown that detectable X-ray cavities should surround the radio lobes. 
A three-dimensional  simulation by \citet{1997MNRAS.284..981C} of a straight light jet was used to model Cygnus\,A. Background subtracted integrations for various energies were presented assuming jet motion in the sky plane. A number of two-dimensional computations by \citet{2013MNRAS.430..174H} of very high Mach number jets confirmed the cavity-shell structures for jets in the sky plane. The X-ray cavities associated with a pair of full magnetohydrodynamic  simulations were presented by  \citep{2011ApJ...730..100M}.
Although full two-sided radio galaxies with evolving jet powers were considered in this work, no asymmetries resulted due to the imposed axisymmetric initial conditions. 
It was found that cavity age and, therefore, cavity power are sensitive to the accuracy of the 
estimated inclination angle of the jet axis.
Cavity age and power estimates within a 
factor of two of the actual values are possible given 
an accurate inclination angle.  
 However, even more crucial to the cavity shape is the  range of ambient and jet properties. For example, the jet-ambient density contrast is able to generate top-wide, centre-wide and 
 bottom-wide X-ray structures as found in the two-dimensional simulations 
 of  \citet{2015ApJ...803...48G} although only the total  Bremsstrahlung cooling rate was modelled.

X-ray surface brightness maps were also generated  by \citet{2020MNRAS.tmp..654T}. 
However, these were based on semi-analytical calculations for  mock FR\,II radio galaxies, enabled 
after several assumptions including self-similarity and axial symmetry. Although a crude model, this work 
reminds us that the X-ray emission from 
distant radio galaxies should become 
dominated by inverse Compton emission off background photons at high redshifts.

We are now able to perform high resolution three dimensional simulations with sufficient efficiency to systematically investigate the influence of jet dynamics and geometry. 
This approach will assume a uniform intracluster medium within which the driving galactic nucleus is stationary.  We will here  study the large-scale X-ray cavities and 
shocks/edges. We mainly analyse the distribution of the hot gas 
through thermal Bremsstrahlung radiation but neglect the internal lobe
X-rays generated from synchrotron,  synchrotron self-Compton.and 
inverse Compton processes.  
Free-free or Bremsstrahlung radiation generally has a low emissivity and so is associated with the large quantities of hot gas in cooling flows
\citep{1974A&A....31..223J}.

Synchrotron emission at X-ray energies is detected from within radio jets and hot spots where strong particle acceleration occurs across shock fronts \citep[e.g.][]{2004ApJ...613..151A}.

Inverse Compton emission is also deduced from within extended radio jets. In this case, the supply of photons to be upscattered can be produced internally 
via synchrotron emission, hence the  term synchrotron  self-Compton. Alternatively, the photons can be  from the cosmic microwave background, galaxy starlight or radiation from the active nucleus 
\citep[e.g.][]{2002ApJ...568..133W, 2005ApJ...622..797K}.
 
We  will here also investigate the ability of the radio galaxy to disrupt a cluster cooling flow. To achieve this, we determine the  speed of advance of the shock wave that propagates into the cluster environment. Besides the Mach number of the advancing bow shock, it is more relevant to provide the Mach number associated with the temperature jump since that is the observable quantity.
For example,  \citet{2011ApJ...734L..28C}  report evidence for a large spheroidal shock in the cluster surrounding the powerful radio galaxy 3C\,444. A temperature jump of approximately1.7 corresponds to a Mach number of 1.7 for a steady hydrodynamic shock.  However, \citet{2011ApJ...734L..28C} remark that
this is likely to be an underestimate since an  average over a large region is made. 

\begin{figure}
\includegraphics[width=0.5\textwidth]{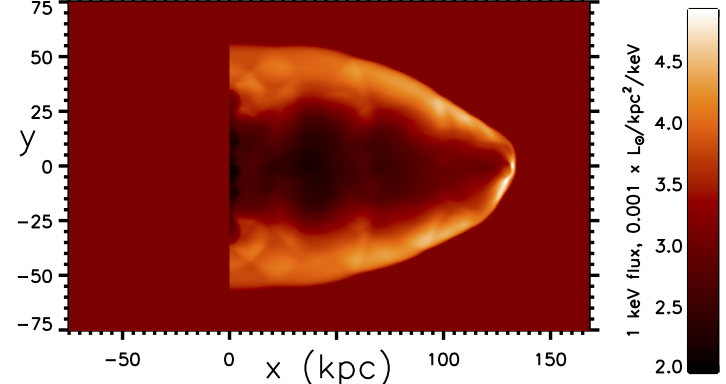}
\includegraphics[width=0.5\textwidth]{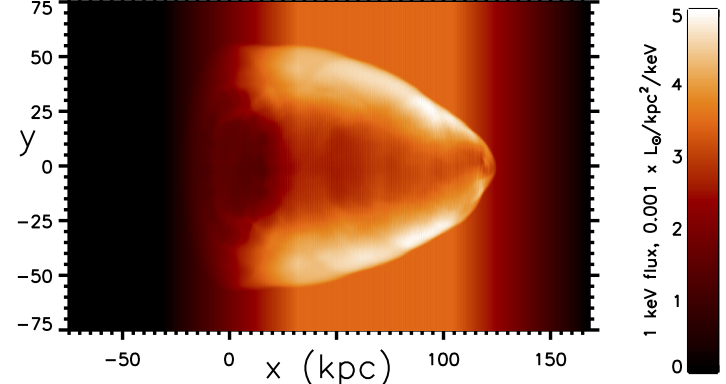}
\caption{  X-ray images generated from the data cube for Model ZDA without the uniform background subtraction. The 1 keV emission is displayed for the jet axis in the sky plane (top panel) and 25$^\circ$ out of the plane (lower panel) for jet-ambient density ratio $\eta = 0.1$, Mach 6 straight jet with 1$^\circ$ precession angle.  This is the free-free X-ray surface brightness; all other images in this paper are sky/background subtracted.  }
\label{ZDA-no-028-00}
\end{figure}
\begin{figure}
\includegraphics[width=0.5\textwidth]{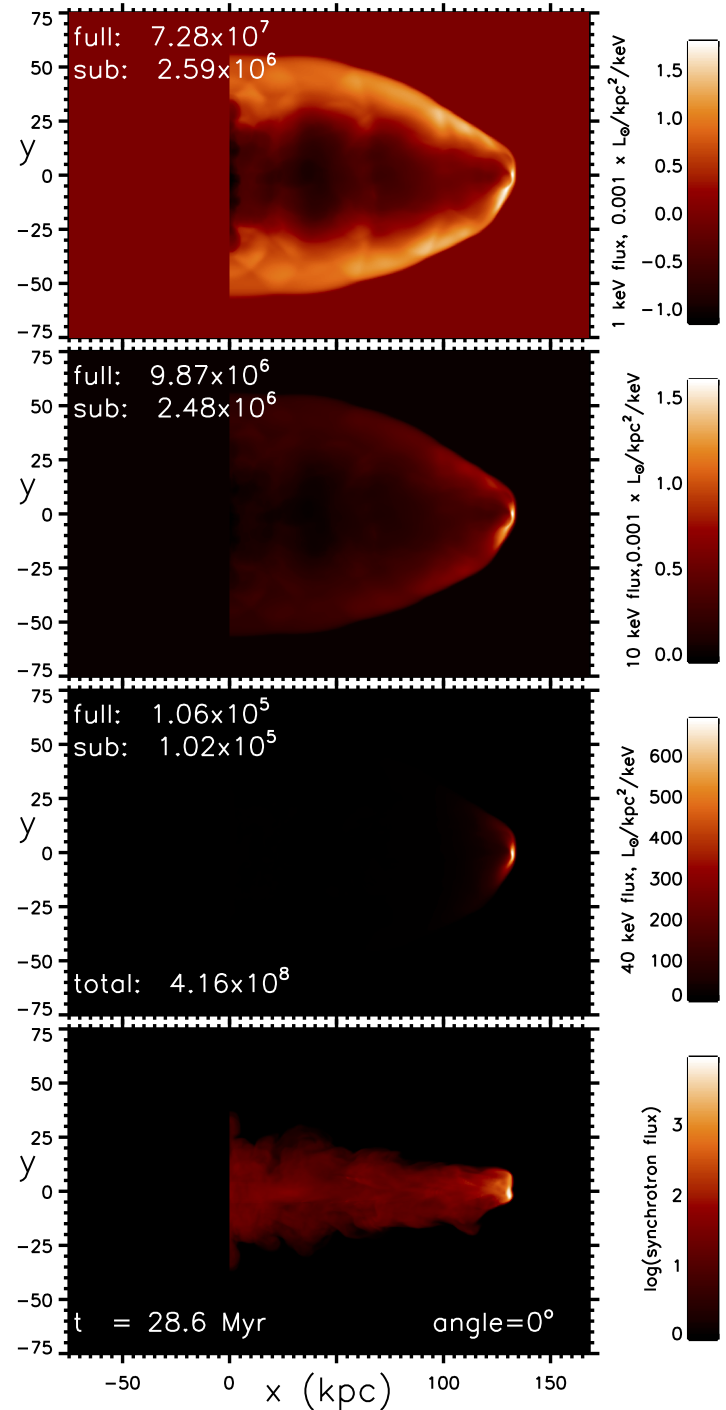}
\caption{ {\bf Background subtracted images for $\eta = 0.1$, Mach 6  jet with 1$^\circ$ precession angle for Model ZDA.}  The top three panels display the free-free X-ray surface brightness at 1\,keV, 10\,keV and 40\,keV   while the lower panel displays a radio image based on a simple synchrotron model. The jet axis is in the plane of the sky.}
\label{ZDA-yes-028-00}
\end{figure}

\begin{figure}
\includegraphics[width=0.5\textwidth]{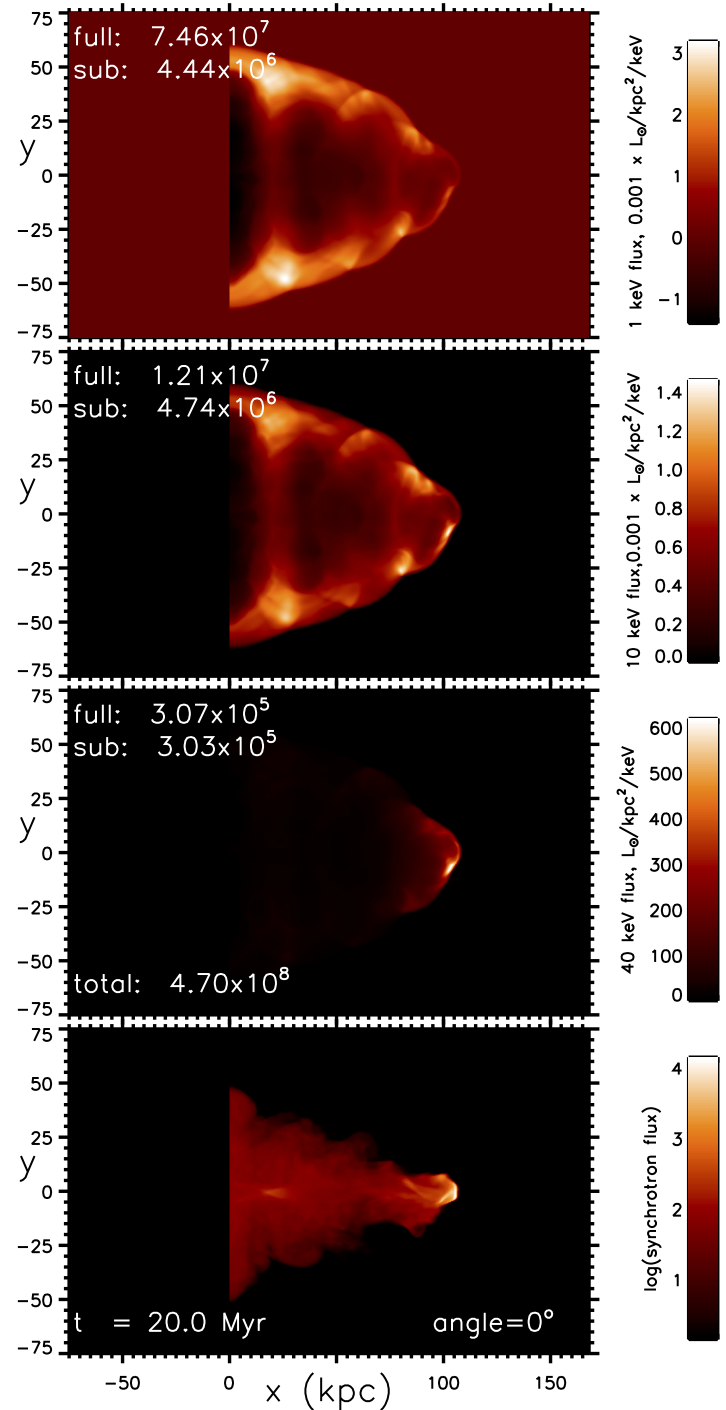}
\caption{ {\bf Background subtracted images for $\eta = 0.01$, Mach 6  jet with 1$^\circ$ precession angle (for Model ZDB).}  The top three panels display the free-free X-ray surface brightness at 1\,keV, 10\,keV and 40\,keV   while the lower panel displays a radio image based on a simple synchrotron model. The jet axis is in the plane of the sky .}
\label{ZDB-yes-020-00}
\end{figure}

\begin{figure}
\includegraphics[width=0.5\textwidth]{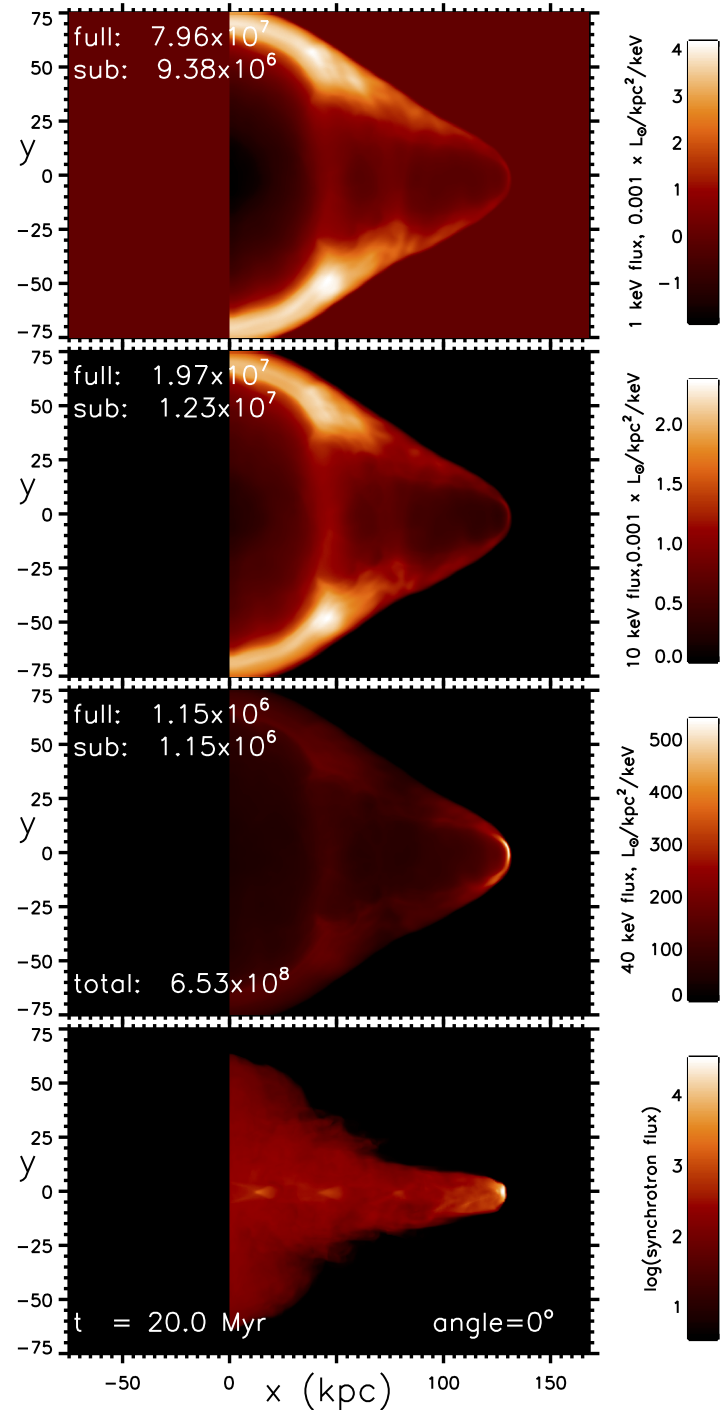}
\caption{ {\bf Background subtracted images for $\eta = 0.001$, Mach 6  jet with 1$^\circ$ precession angle (for Model ZDC).}  The top three panels display the free-free X-ray surface brightness at 1\,keV, 10\,keV and 40\,keV   while the lower panel displays a radio image based on a simple synchrotron model. The jet axis is in the plane of the sky.}
\label{ZDC-yes-020-00}
\end{figure}
\begin{figure}
\includegraphics[width=0.5\textwidth]{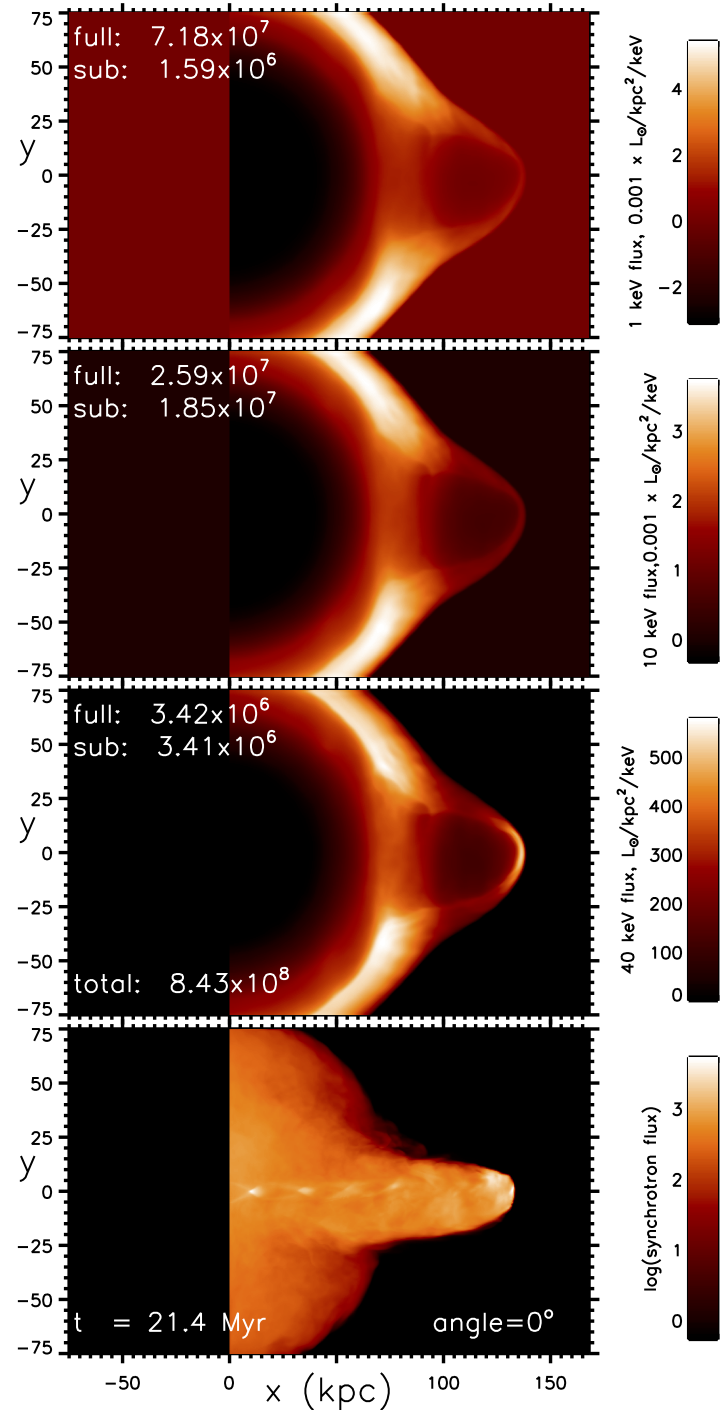}
\caption{ {\bf Background subtracted images for a very light $\eta = 0.0001$, Mach 6  jet with 1$^\circ$ precession angle(for Model ZDM).}  The top three panels display the free-free X-ray surface brightness at 1\,keV, 10\,keV and 40\,keV   while the lower panel displays a radio image based on a simple synchrotron model. The jet axis is in the plane of the sky .}
\label{ZDM-yes-021-00}
\end{figure}

\begin{figure}
\includegraphics[width=0.5\textwidth]{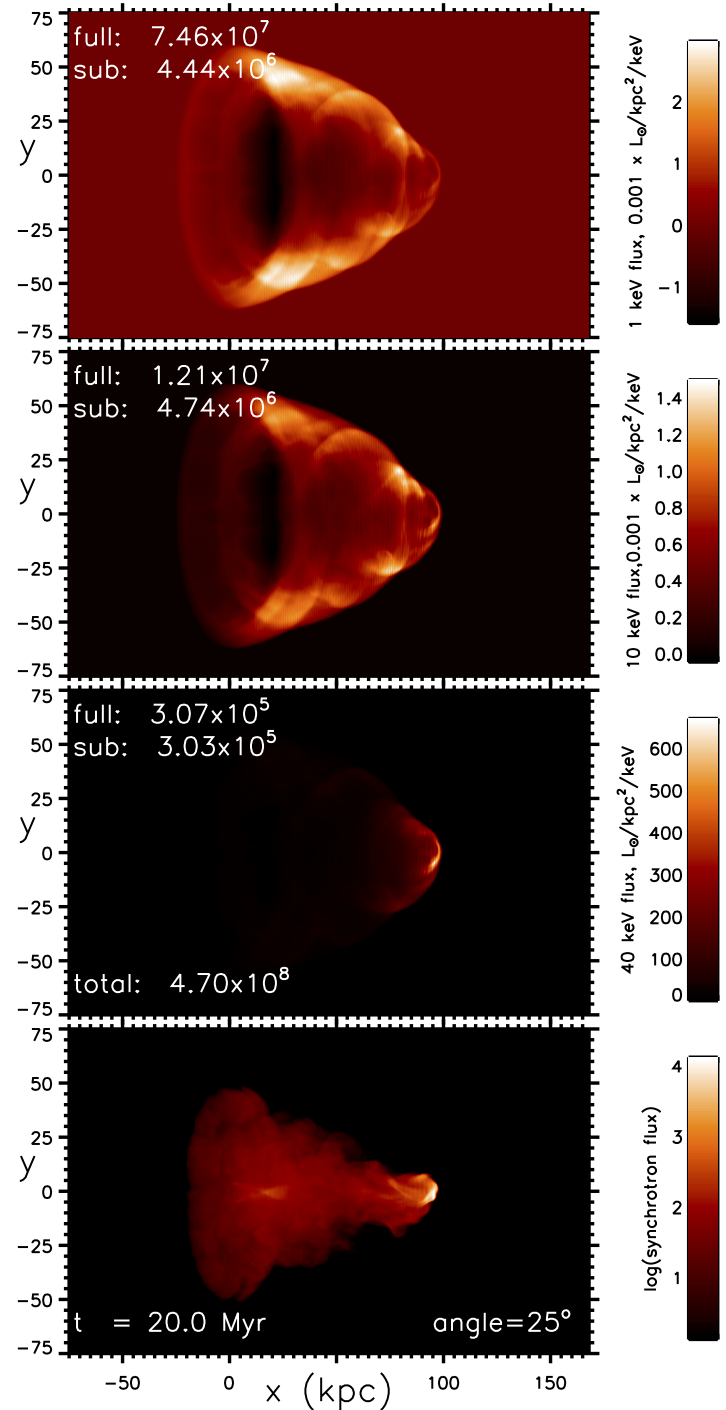}
\caption{ {\bf The Mach 6 straight jet with the jet axis  at an angle of
  25$^\circ$  to the plane of the sky.}   These are background subtracted images produced by a  jet with a jet-ambient density ratio of 0.01 and a slow precession with half-angle of
  1$^\circ$ (for Model ZDB). The top three panels display the free-free X-ray surface brightness at 1\,keV, 10\,keV and 40\,keV   while the lower panel displays a radio image based on a simple synchrotron model. }
\label{ZDB-yes-020-25}\end{figure}

\begin{figure}
\includegraphics[width=0.5\textwidth]{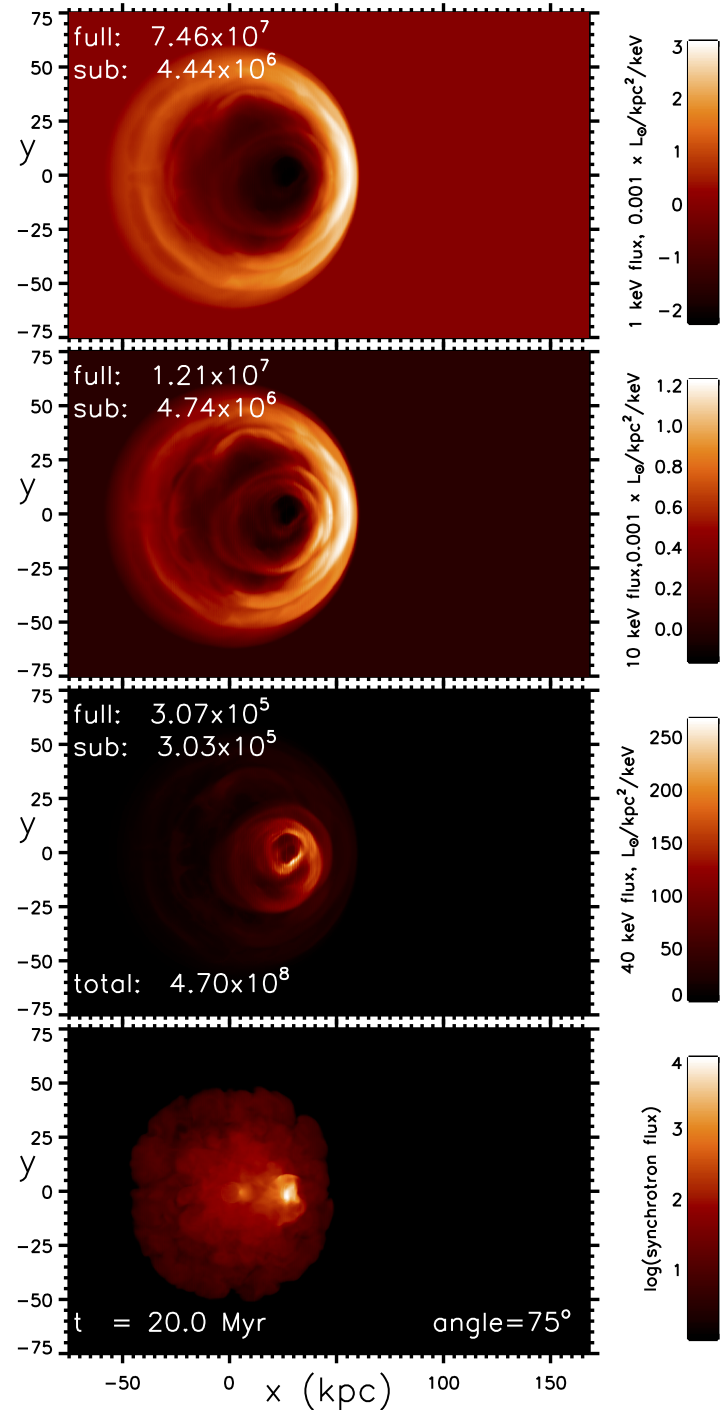}
\caption{ {\bf The Mach 6 straight jet with the jet axis  at an angle of
  75$^\circ$  to the plane of the sky.}   These are background subtracted images produced by a  jet with a jet-ambient density ratio of 0.01 and a slow precession with half-angle of
  1$^\circ$ (for Model ZDB). The top three panels display the free-free X-ray surface brightness at 1\,keV, 10\,keV and 40\,keV   while the lower panel displays a radio image based on a simple synchrotron model.}
\label{ZDB-yes-020-75}
\end{figure}

\begin{figure}
\includegraphics[width=0.5\textwidth]{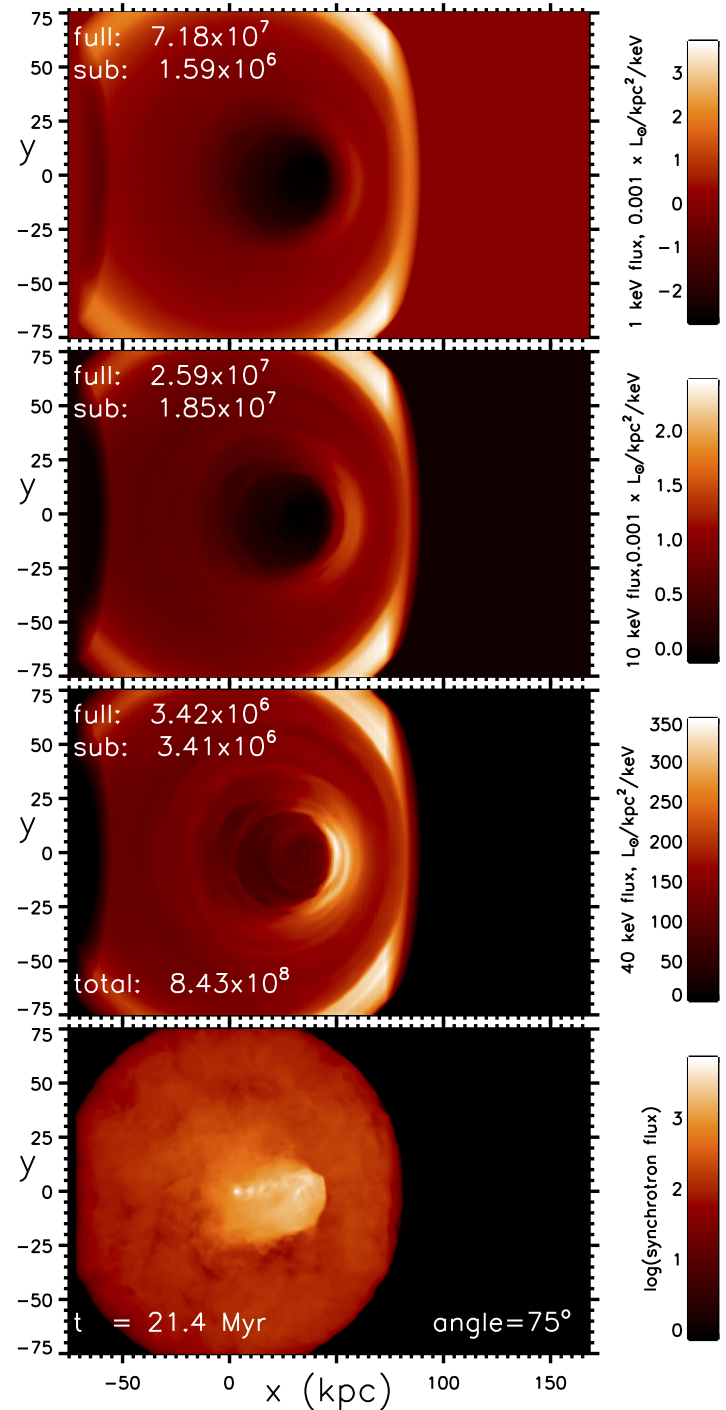}
\caption{ {\bf The $\eta = 0.0001$ Mach 6 straight jet with the jet axis  at an angle of
  75$^\circ$  to the plane of the sky.}   These are background subtracted images produced by a  jet with a jet-ambient density ratio of 0.0001 and a slow precession with half-angle of
  1$^\circ$ (for Model ZDM). The top three panels display the free-free X-ray surface brightness at 1\,keV, 10\,keV and 40\,keV   while the lower panel displays a radio image based on a simple synchrotron model.}
\label{ZDM-yes-021-75}
\end{figure}

\begin{figure}
\includegraphics[width=0.5\textwidth]{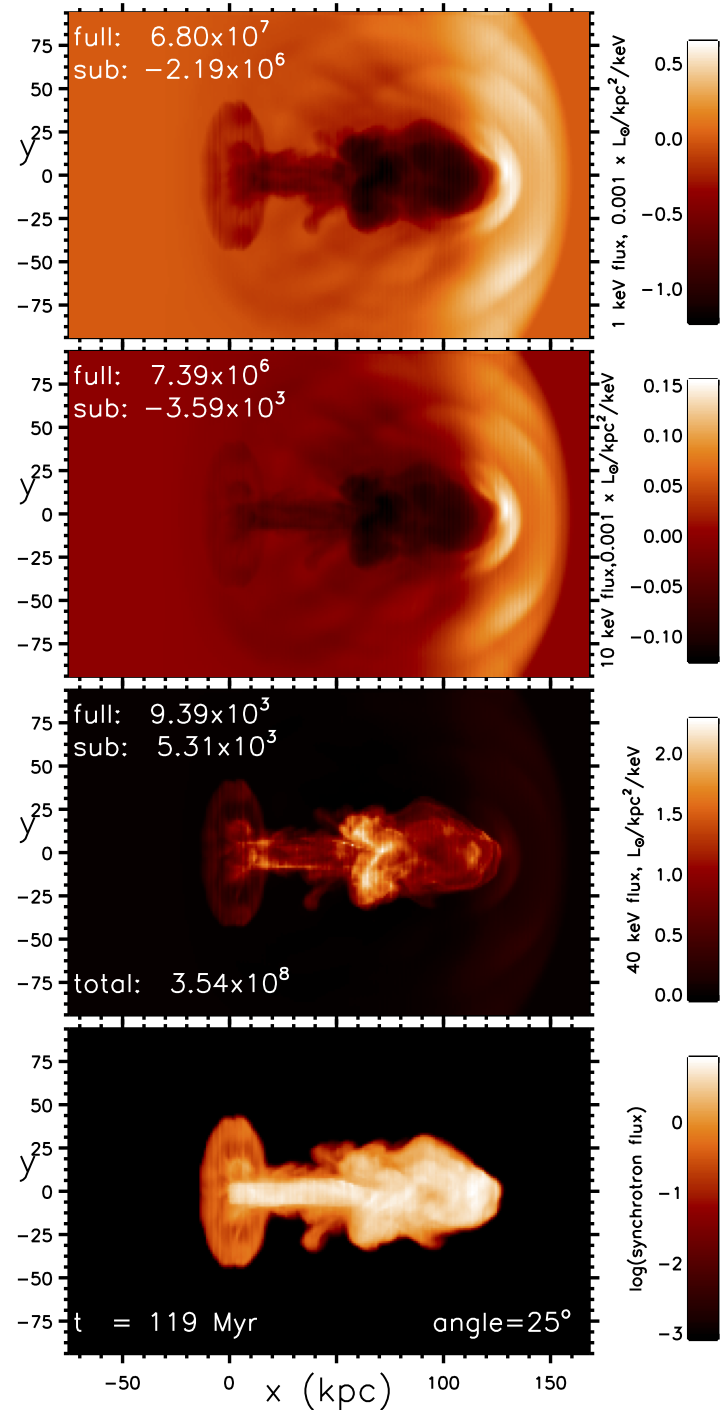}
\caption{ {\bf A Mach 2 straight jet (fModel M2), viewing angle 25$^\circ$ and $\eta = 0.1$ with 1$^\circ$ precession.}   Background subtracted images. The top three panels display the free-free X-ray surface brightness at 1\,keV, 10\,keV and 40\,keV   while the lower panel displays a radio image based on a simple synchrotron model. The jet axis is at an angle of
  25$^\circ$  to the plane of the sky. The length scales are expressed in kiloparsecs.}
\label{M2-yes-119-25}
\end{figure}

\begin{figure}
\includegraphics[width=0.5\textwidth]{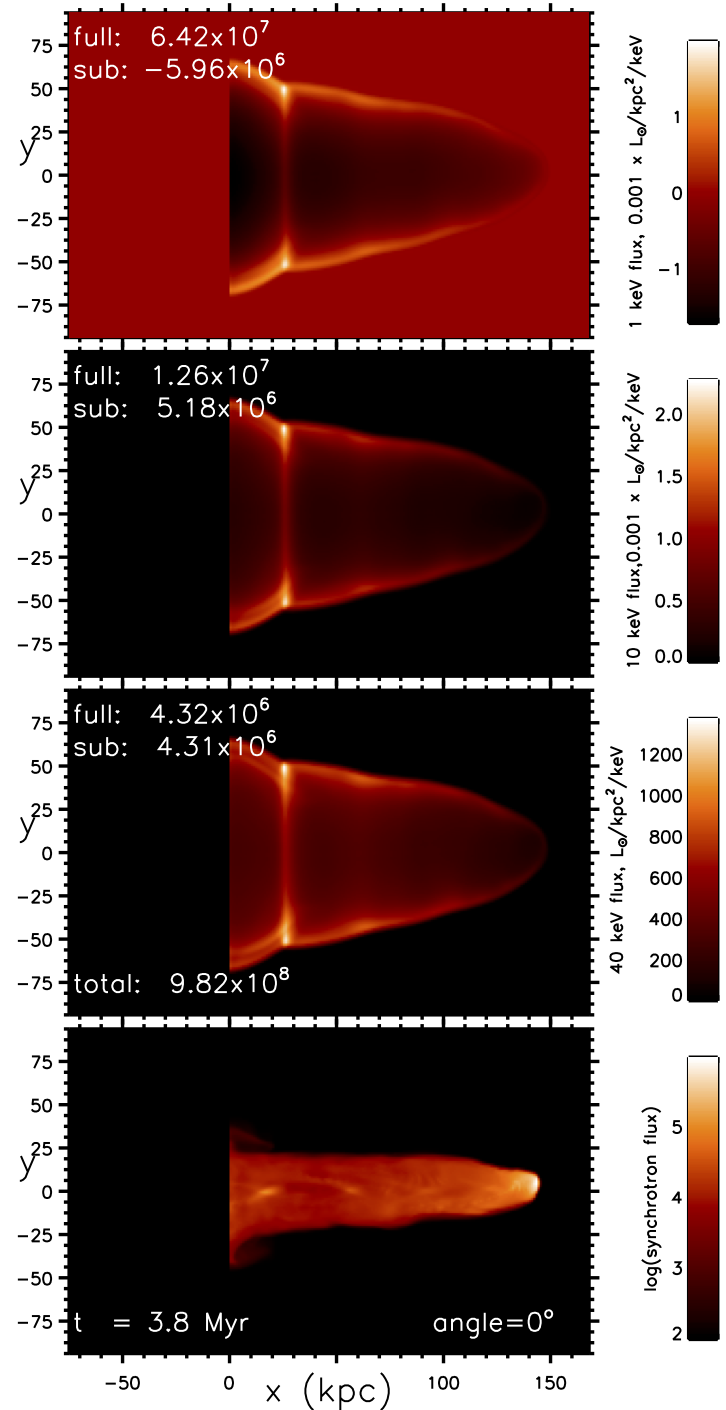}
\caption{ {\bf A Mach 24 straight jet, viewing angle 0$^\circ$ and $\eta = 0.1$ (for Model IC).}   Background subtracted images. The top three panels display the free-free X-ray surface brightness at 1\,keV, 10\,keV and 40\,keV   while the lower panel displays a radio image based on a simple synchrotron model. The jet axis is in the plane of the sky. }
\label{IC-yes-003-00}
\end{figure}
\begin{figure}
\includegraphics[width=0.5\textwidth]{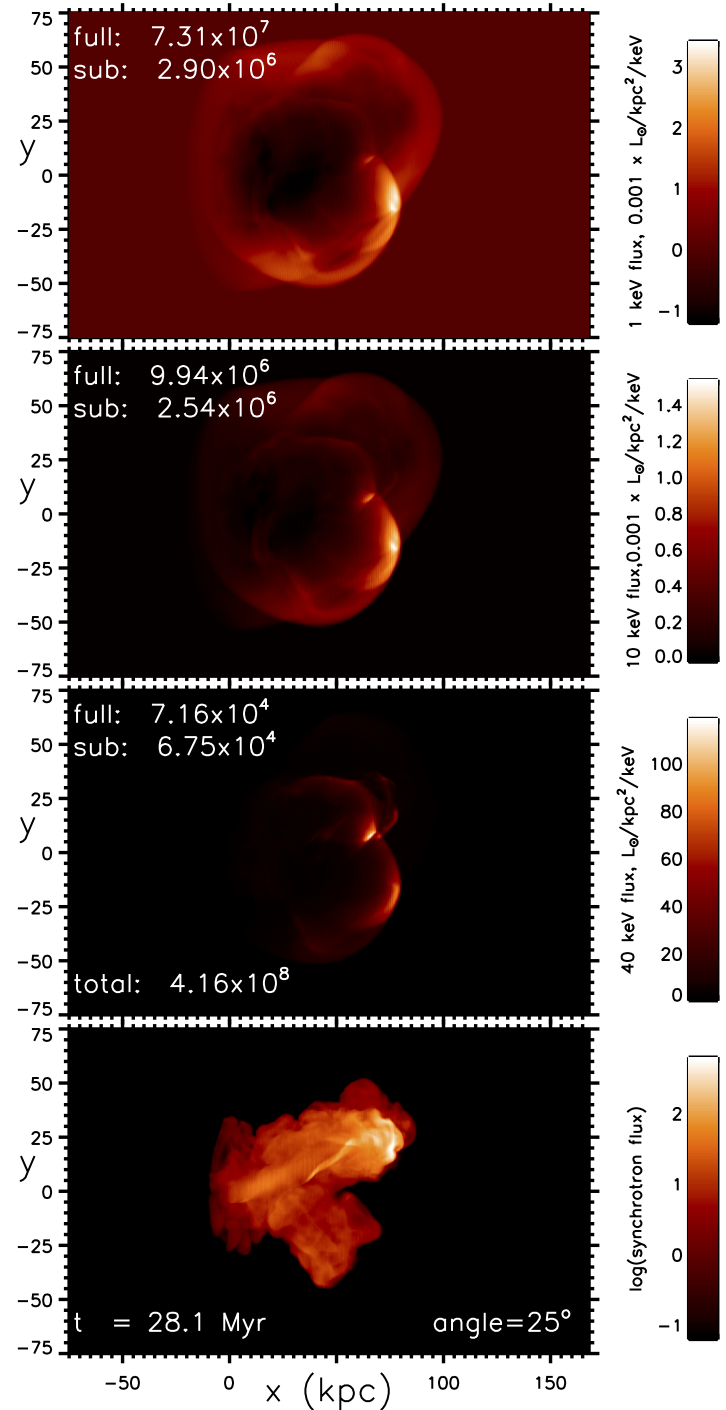}
\caption{ {\bf A slow precessing jet with 20$^\circ$ precession angle (for Model ZDG). image is from the projected X-Y plane}.   Images produced by a  Mach 6 jet with a jet-ambient density ratio of 0.1. The top three panels display the free-free X-ray surface brightness at 1\,keV, 10\,keV and 40\,keV   while the lower panel displays a radio image based on a simple synchrotron model. The jet axis is at an angle of
  25$^\circ$  to  the plane of the sky.}
\label{ZDG-yes-028-25y}
\end{figure}

\begin{figure}
\includegraphics[width=0.5\textwidth]{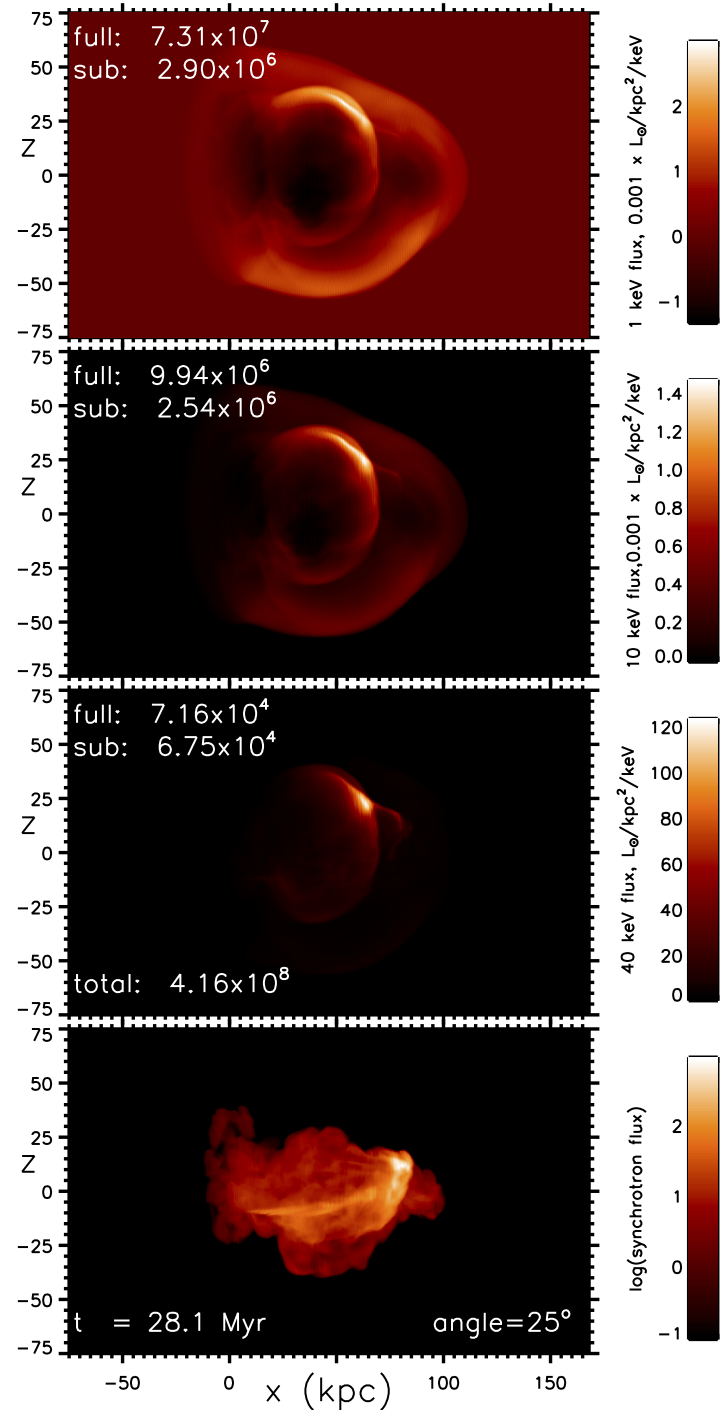}
\caption{ {\bf  A slow precessing jet with 20$^\circ$ precession angle. image is from the projected X-Z plane}.   Images produced by a  Mach 6 jet with a jet-ambient density ratio of 0.1 (for Model ZDG). The top three panels display the free-free X-ray surface brightness at 1\,keV, 10\,keV and 40\,keV   while the lower panel displays a radio image based on a simple synchrotron model. The jet axis is at an angle of
  25$^\circ$  to  the plane of the sky.}
\label{ZDG-yes-028-25z}
\end{figure}

\begin{table*}
\resizebox{0.9\textwidth}{!}{\begin{minipage}{\textwidth}
  \centering
  \caption{Summary of jet properties as expressed for a domain scaled to 150\,kpc. The run identification code is consistent with previous papers in this series with the label 
 Z denoting a reflection boundary condition on the injection plane outside the nozzle and the 
 D corresponding to larger 300$^{3}$ domains.}
    \begin{tabular}{crrlrrr}
    \hline
 Figure  & Code used & Jet Mach & Density     & Prec.     & Jet Sound Speed  & Jet Power   \\          number & in Paper~1  & number   & ratio  $\eta$  & angle & cm\,s$^{-1}$      &       erg\,s$^{-1}$\\
    \hline
\ref{ZDA-no-028-00}/\ref{ZDA-yes-028-00} & ZDA & 6  & 0.1  & 1$^\circ$ &  3.24$\times$10$^8$  & 6.42$\times$10$^{44}$    \\
 \ref{ZDB-yes-020-00}   & ZDB  & 6   & 0.01  &  1$^\circ$   & 
1.02$\times$10$^9$  & 2.03$\times$10$^{45}$    \\
\ref{ZDC-yes-020-00}  & ZDC   & 6    & 0.001    &   1$^\circ$ &   
3.24$\times$10$^9$  & 6.42$\times$10$^{45}$    \\
\ref{ZDM-yes-021-00}   & ZDM   & 6    & 0.0001  &   1$^\circ$ &
1.02$\times$10$^{10}$  & 2.03$\times$10$^{46}$  \\
\ref{ZDB-yes-020-25}  & ZDB  & 6   & 0.01   &  1$^\circ$ &  3.24$\times$10$^8$  & 6.42$\times$10$^{44}$    \\
\ref{ZDB-yes-020-75}  & ZDB  & 6    & 0.01    & 1$^\circ$ &  3.24$\times$10$^8$  & 6.42$\times$10$^{44}$    \\
\ref{ZDM-yes-021-75}  & ZDM & 6  & 0.0001  &   1$^\circ$ &   1.02$\times$10$^{10}$  & 2.03$\times$10$^{46}$  \\
 \ref{M2-yes-119-25}     & M2   & 2   & 0.1     &   1$^\circ$      &  3.24$\times$10$^8$  & 2.38$\times$10$^{43}$   \\  
  \ref{IC-yes-003-00}    & IC      &24   & 0.1  &   1$^\circ$ &
1.02$\times$10$^9$  & 1.30$\times$10$^{47}$    \\

  \ref{ZDG-yes-028-25y}/\ref{ZDG-yes-028-25z} & ZDG  & 6 & 0.1  &  20$^\circ$  & 3.24$\times$10$^8$  & 6.42$\times$10$^{44}$    \\
     \ref{ZDG4-yes-047-50z}-xz & ZDG4  & 6  & 0.1  & 20$^\circ$\,fast & 3.24$\times$10$^8$  & 6.42$\times$10$^{44}$    \\
 \\
     \hline
     \end{tabular}%
  \label{tab:parameters}%
\end{minipage} }
\end{table*}%


\section{Method} 
\label{Method}


\subsection{Simulation  and display parameters}

The simulations analysed here have been introduced in Paper\,1 \citep{2016MNRAS.458..558D} for their physical attributes and in  Paper\,2 \citep{2019MNRAS.490.1363S}.   for their   radio morphology and classification. The set of defining parameters and corresponding figures are provided in  Table\,\ref {tab:parameters}.

The analysis was performed with algorithms incorporated into IDL software.
Before inspecting the figures, some common properties should be noted in advance (in order to maximise the panel sizes and reduce the caption lengths). The length scales are expressed in kiloparsecs although the pixel unit is 0.5\, kiloparsecs which corresponds to the zone sizes of the fixed-grid Eulerian {\em Pluto} code. Luminosities are in solar units with the total Bremsstrahlung cooling (total), full emission from the specific band per keV (full) , and subtracted surface brightness (sub) where the background has been subtracted (not subtracted for the first figure). See Table\,\ref{tab:model} for a summary.

The simulations were performed with PLUTO, a grid-based code incorporating  modern Godunov-type shock-capturing schemes \citep{2007ApJS..170..228M}. After comparing the results of  numerous options, we chose a fast linear interpolation  time-stepping (denoted HLLC) scheme as developed by Harten, 
Lax and Van Leer and detailed by \citet{1994ShWav...4...25T}.
The simulations were mainly performed on grids of 300$^3$ unless otherwise stated. Hence the voxel and pixel sizes shown in the figures corresponds to  0.5\,kpc.

We have performed a resolution study with grids of 75$^3$, 150$^3$,  225$^3$ and 300$^3$. We presented results in Paper 1 that showed there was a gradual convergence with  150$^3$ sufficient for most purposes and  225$^3$  oroviding reliable hot spot properties.

\subsection{Parameters for the ambient medium}

The temperature of the environment is taken as $T = 4.56 \times 10^7$~K, corresponding to 4 keV, with a hydrogen nuclei density of $n = 1.21 \times 10^{-3}$~cm$^{-3}$
to simulate one half of a Radio Galaxy within a cube which stretches  150~kpc. This is adjustable to other densities and length scales quite easily since
 the flows are adiabatic with a specific heat ratio of 5/3. It would also be possible to scale the derived emission provided  the ratio of the X-ray energy to the temperature were held fixed.  
 
 A uniform ambient medium is taken as opposed to the density profiles commonly assumed to model galactic atmospheres such as by  \citet{1997MNRAS.284..981C}.
 We model the intracluster medium in this instance but realise that cluster cooling flows and galactic motions would need us to  introduce detailed structure.   
\subsection{Parameters for the jet}

We configured the  code for  non-relativistic hydrodynamics.  We do this in order to explore ranges in the dynamics in full three-dimensions with a reasonable resolution.  Full details were published in Paper\,1 for the physical set up and in Paper\,2 for the radio emission.
 
 The jet is initially in pressure balance with the ambient medium but  the sound speed in the jet is considerably higher given the low density jet which is expressed by the  density ratio $\eta$.  It is important to allow the jet direction to vary, not only to break symmetry but also because there is considerable evidence for this \citep{2019MNRAS.482..240K}.Therefore, the  jet direction is precessed through sinusoidal velocity changes in the y and z directions 
 as detailed in \citet{2016MNRAS.458..558D}. The standard slow precession rate has an angular frequency of $\pi$/2 in units of $c_s/r_j$ where $c_s$ is the ambient sound speed and $r_j$ is the initial jet radius. This corresponds to a precession period  of  19.1\,Myr with the present units.

\subsection{Parameters for the emission}

 To calculate the total Bremsstrahlung emission from the simulations, we assume full ionisation, with 10\% helium by number,.
 From the theoretical foundations  \citep{1975rpa..book.....T, 1979rpa..book.....R}, there are a number of\
 approximations fo the formula summarised by \citet{2020ApJ...898...50Y} from which we employ  
 \begin{equation}\label{eq_brem1}
      L_\text{brem} =  1.42 \times 10^{-27} \cdot n_e^2  \cdot  T^{1/2}  \text{erg}~\text{s}^{-1}~\text{cm}^{-3} ,
 \end{equation}
 where $T$ is the temperature in Kelvin and $n_e$ is the electron density measured in cm$^{-3}$.

We present emission maps at soft kT$_o$=1\,keV ($2.42 \times  10^{17}$~Hz), intermediate  kT$_o$ = 10\,keV and hard  kT$_o$ = 40\,keV X-rays. The formula employed is
      \begin{equation}\label{eq_brem2}
       L_\text{o} =  6.8 \times 10^{-38} \cdot n_e^2  \cdot  T^{-1/2} \cdot exp(-T_o/T)  \text{erg}~\text{s}^{-1}~\text{cm}^{-3}~\text{Hz}^{-1}.
\end{equation}

Note that we will consider the emission at three specified energies but multiply up to band passes of width  1\,keV and provide values per square kiloparsec which generates meaningful luminosities. Scales are provided on  bars but are adjusted by appropriate factors as indicated in their titles for display purposes.

\section{Results} 

\subsection{Density dependence }

 The density of the jet relative to the ambient medium is shown to have a major influence on the X-ray distribution. We define $\eta$ as the jet-ambient density ratio and consider light jets with $\eta$ between 0.1 and 0.0001.
 
 With $\eta = 0.1$, a substantial depression in the soft X-rays is revealed as displayed in the top panel of Fig.\,\ref{ZDA-no-028-00}. This particular figure is not background subtracted and the data set is confined to within a 150\,kpc\,$\times$\,150\,kpc region filled with 300$^2$ pixels. The display has been chosen so that when the 300$^3$ data cube is rotated (see below) , the image will remain in the window with the origin of the jet remaining at a fixed point.
 
Probably more instructive are the  equivalent background subtracted  images displayed in Fig.\,\ref{ZDA-yes-028-00}.  The central cavity is enveloped by a thick elliptical shell. As expected, the highest surface brightness coincides with the leading edge where the heating and compression are strongest. However,  the total emission is dominated by the extended flanks which consist of locally shocked gas and adiabatically-expanding hot gas flowing within.

This shell is much dimmer at intermediate X-ray energies shown in the second panel while this is then reduced to an arc at hard X-rays in the third panel. It is clear that the hard X-ray energy falls well into the exponential part of the emission function for all except the very hot leading part of the bow.  In addition, the elliptical shell is distorted with stronger emission to the 'south' in all bands.
 
A different structure is apparent for $\eta = 0.01$ shown in Fig.\, \ref{ZDB-yes-020-00}. Firstly, the shell is better described as conical when the jet axis lies in the sky plane. Ribs of emission are now very prominent, running across the shell perpendicular to the jet axis. Four ribs can be seen at the stage shown. The separation between is approximately 60\,kpc. The nature of the ribs will be elucidated below through viewing the structure at another angle.

At this lower density, the leading shock is stronger and wider. The jet has penetrated across the grid in a shorter time of 20\,Myr, displaying a streamlined structure rather than the blunt appearance of the $\eta = 0.1$ case above.

With the jet density further reduced, as shown in Fig.\,\ref{ZDC-yes-020-00}  and \ref{ZDM-yes-021-00}, the ribs disappear. Instead, there is a prominent  
inner cavity in soft X-rays, which is roughly oblate. This is enshrouded by bright shoulders. The visible knotty radio jet now penetrates through the inner cavity although the hot cocoon immediately surrounding it does not seem very distinct. At the lowest density ratio, the entire disturbed region contains two cavities: the inner oblate at all energies and an outer  circular bubble at intermediate and hard energies. 
 
\subsection{Orientation of the jet axis }
 
 The  integrated line of sight emission with the jet axis at various angles is presented with the panel widths having been chosen so that the entire projected data cube is visible and the jet origin remains at a fixed location, the origin (0,\,0).
 
The X-ray structure is altered systematically as the orientation out of the plane of the sky is increased. This effect is illustrated in 
 Figs.\,\ref{ZDB-yes-020-25} at    25\degsy  and  
 Fig.\,\ref{ZDB-yes-020-75}  at    75\degsy  .
 to the plane of the sky.\\At soft X-ray energies, the cone turns into a spindle and the ribs become round ripples. At progressively high energies, the deep cavity becomes smaller and shifts towards the front. Note also that secondary radio hotspots are present at these angles due to projection effects of filamentary structures.
 
 A remarkable short knotty radio jet is revealed in the lower panel of Fig.\,\ref{ZDM-yes-021-75} for the lowest density jet at a low inclination to the line of sight. This jet is surrounded by an extended radio halo. Moreover, a similar sized thick X-ray front exists at the edge of the halo. Therefore, this simulation may have relevance to a number of quasar jets with halo properties. 
 
\subsection{Jet Mach number } 
 
 At low Mach numbers, the radio galaxy displays considerable turbulent structure as shown in the lower panel of 
 Fig.\,\ref{M2-yes-119-25}. Although on these scales the jet has not completely disintegrated, there is no advancing hotspot. The soft X-ray cavity precisely follows the radio lobe. Ahead of the lobe is a very wide and thick shell of enhanced thermal emission. Hence, low Mach number jets can energise a large volume of the intracluster medium.

 In hard X-rays, the structure is reversed with only the very hottest regions contributing. This includes a sheath of hot gas surrounding the jet out to 60 kiloparsec, detected here as two thin parallel filaments. This is followed by a turbulent zone coinciding with the jet disruption. Hence, Bremsstrahlung emission can arise internal to the radio lobes at low Mach numbers, partly due to the stronger dissipation within the radio galaxy and weaker transmission to the intracluster medium (see Column 11 of Table\,\ref{tab:model}.

 At high Mach numbers, a contrasting pattern emerges. As shown in Fig.\,\ref{IC-yes-003-00} for a Mach number of 24 and $\eta = 0.001$, a  thimble-like cavity is commensurate with the streamlining due to the high Mach number.   This is consistent with the short grid crossing time of 7\,Myr. Note that the injected jet is pressure-matched to the environment. That implies that a low-density jet will possess a high sound speed. This compensates in terms of the momentum flow for a given Mach number. Here, however, the advanced speed should be proportional to the Mach number, as indeed found.

\subsection{Precession angle } 
 
 The Mach numbers for shocks propagating ahead of radio galaxies have been estimated through their X-ray properties. The shocks tend to be very weak with Mach numbers below 3. Moreover, the radio lobes are generally not greatly streamlined. Together, these facts suggest that the speed of propagation into the intracluster medium is only mildly supersonic despite the very high Mach numbers of the driving jets.  One way to achieve this in the present context is through a slow precession of the jet axis so that the jet momentum is distributed over a wide surface area. 
 
 Precession can manifest itself as multiple radio sub-lobes as shown in Fig.\,\ref{ZDG-yes-028-25y}, reminiscent of X-shaped or double-double radio galaxies. In X-rays, this simulation has a unique signature: the high surface brightness is associated with the relic radio component rather than with the present hotspot. This holds for the soft and intermediate X-ray energies but the newly advancing lobe is more apparent at hard X-rays. Hence, the Bremsstrahlung emission can remain  dominated by the large amount of swept-up and compressed material from the previously compressed environment while the jet rotates and progresses through an earlier ejection of lobe material. 
 
 Rotated about the jet axis by 90$^\circ$, yields a contrasting radio morphology as displayed in the lower panel of Fig.\,\ref{ZDG-yes-028-25z}. The jet begins quite straight 
 snd becomes slightly curved before deflection at a secondary hotspot. In this case, the peak  soft X-ray emission does not coincide with either hotspot but appears where the overspill from the primary hotspot has created a new lobe to the top of the image. A further thick rim is present on the opposite side.
 
 On comparison of the two images, it is apparent that there are two circular cavities. The smaller inner cavity corresponds to the region associated with the relic lobe while the large cavity is associated with the overall evacuated lobe. 
 
\subsection{Luminosity \& energetics}

The total Bremsstrahlung luminosity is indicated directly on the third panel of the figures. In the absence of the radio galaxy, the luminosity from the 150\,kpc cube is 3.61 $\times$\,10$^8$\,L$_\odot$. It should be noted that scaling to a sphere of radius 150\,kpc with a hydrogen nuclei density of 3\,$\times$\,10$^{-3}$\,cm$^{-3}$ would yield an X-ray luminosity of 1.35 $\times$\,10$^{10}$\,L$_\odot$.   Cooling times remain longer than the age of the Universe.
  
  The introduction of the jet increases the total luminosity in all cases despite the cavity formation.  The cavity forms through  the  compression of material. and the shell or rim  is generally responsible for enhancing the Bremsstrahlung emissivity. The one exception is the Mach 2 jet in which significant amounts of ambient material have been expelled from the grid.
  
The total luminosity increases  as the jet density is reduced but not by a large factor, approximately doubling over the entire density range of a 1,000. This reflects the fact that a lower density is matched by a higher jet sound speed,
which maintains the same momentum flow rate and approximately the same  advance speed of the radio hotspot and interface. 

Also shown in each panel are the full Bremsstrahlung fluxes in solar units per keV. Underneath this, the 
background subtracted fluxes are given. This resets the zero point on the colour bar, allowing the cavities to be emphasized as negative values. In this respect, the soft X-ray cavity does deepen as the jet density decreases while, on the other hand,  the shell surface brightness increases. Overall, upon background subtraction, the 
net integrated flux combining   enhanced and reduced regions is relatively small. It is also clear that cavities are prevalent in soft X-rays but not in hard X-rays where the background is almost 
negligible and the emission is mainly generated from  the leading edge of the ambient  bow shock.

The size and depth of the cavities are quantified and presented in Column 9 of  Table\,\ref{tab:model}. The radio lobe area is also listed in Column\,8. The detectable lobe is here given a threshold of 1\% of the peak value. Hence the radio lobe and the cavity area both generally increase as the jet-ambient density ratio decreases.
It can also be noted that the radio area and cavity area can greatly depend on the viewing angle. The depth of the X-ray cavity is  measured as the total surface brightness  of the cavity within the area, defined as 0.3 below the background in the figures. It is evident that cavities are deeper when the jet axis is closer to the line of sight. However,  several factors contribute. For example, 
the high Mach number straight jet has blown a deep cavity although the streamlined shape means that it has not vacated a large volume. On the other hand, the low Mach number jet has also cleared a deep cavity coincident with the radio lobe.

\begin{table*}
\resizebox{0.9\textwidth}{!}{\begin{minipage}{\textwidth}
  \centering
  \caption{Properties of X-ray images of simulated radio galaxies.}
    \begin{tabular}{crlrrrrrrrrrr}
    \hline
   Fig. & Mach  & Density        & Prec.   & Angle to    & Interface     & Bow         &  Radio   & Cavity             & Cavity depth     & Energy & X-ray & X-ray \\
          & number     & ratio  $\eta$   & angle  & sky plane  &  Mach No. & Mach No. & area kpc$^2$& area kpc$^2$ &   L$_X/10^6$L$_\odot$    & transfer & Temp.10$^7$K & Mach No. \\
    \hline
\ref{ZDA-no-028-00}/\ref{ZDA-yes-028-00} & 6  & 0.1  & 1$^\circ$ & 0$^\circ$  & 4.35  & 4.48 & 142  &4,592  & -1.99  &  0.71  & 5.23  &  1.12  \\
 \ref{ZDB-yes-020-00}    & 6    & 0.01      &   1$^\circ$ & 0$^\circ$   & 5.22   & 5.42    &   214   &  4,634    & -2.11  &  0.86  & 6.00  & 1.30\\
 \ref{ZDC-yes-020-00}    & 6    & 0.001    &   1$^\circ$ & 0$^\circ$   & 6.10   & 6.29    &   464    & 4,187   & -2.91  &  0.92   & 7.48  & 1.60\\
\ref{ZDM-yes-021-00}     & 6    & 0.0001  &   1$^\circ$ & 0$^\circ$   & 5.89  & 6.18   & 10,292   & 6,940  & -15.78 &  0.94  & 10.25  & 2.09 \\
\ref{ZDB-yes-020-25}      & 6    & 0.01      &   1$^\circ$ & 25$^\circ$ & 5.22  & 5.42  &    380   &  4,533    & -2.42  &  0.87   & 6.00  & 1.30    \\
\ref{ZDB-yes-020-75}      & 6    & 0.01    &   1$^\circ$ & 75$^\circ$  & 5.22  & 5.42    &   543   &  4,575     & -5.00  &  0.87 & 6.00  & 1.30 \\
\ref{ZDM-yes-021-75}      & 6  & 0.0001  &   1$^\circ$ & 75$^\circ$  & 5.89  & 6.18 & 15,785  & 11,745    & -8.03  &  0.94 & 10.25  & 2.09\\
 \ref{M2-yes-119-25}        & 2   & 0.1  &   1$^\circ$      & 25$^\circ$  & 1.07   & 1.19 & 5,740   & 19,348    & -4.34  &  0.59  &  4.71  &  1.01 \\  
  \ref{IC-yes-003-00}          &24   & 0.1  &   1$^\circ$ & 25$^\circ$  & 35.50   & 37.25 & 2,765  & 8,566     & -6.83  &  0.86  & 6.40 & 1.38\\
  \ref{ZDG-yes-028-25y}-xy   & 6 & 0.1  &  20$^\circ$ & 25$^\circ$  & 0.80   & 3.30 & 3,024     & 3,769    & -2.02 &  0.79 & 5.23 & 1.13 \\
  \ref{ZDG-yes-028-25z}-xz     & 6 & 0.1  &  20$^\circ$ & 25$^\circ$  & 0.80   & 3.30 & 1,960     & 3,989    & -1.99 &  0.79 & 5.23 & 1.13 \\ 
\ref{ZDG4-yes-047-50z}-xz   & 6  & 0.1  & 20$^\circ$\,fast & 50$^\circ$  & 1.44  &  2.37 & 7,217  &  6,532  & -5.79 &  0.82  & 5.50  &  1.19  \\
     \hline
     \end{tabular}%
  \label{tab:model}%
\end{minipage} }
\end{table*}%

\begin{figure}
\includegraphics[width=0.5\textwidth]{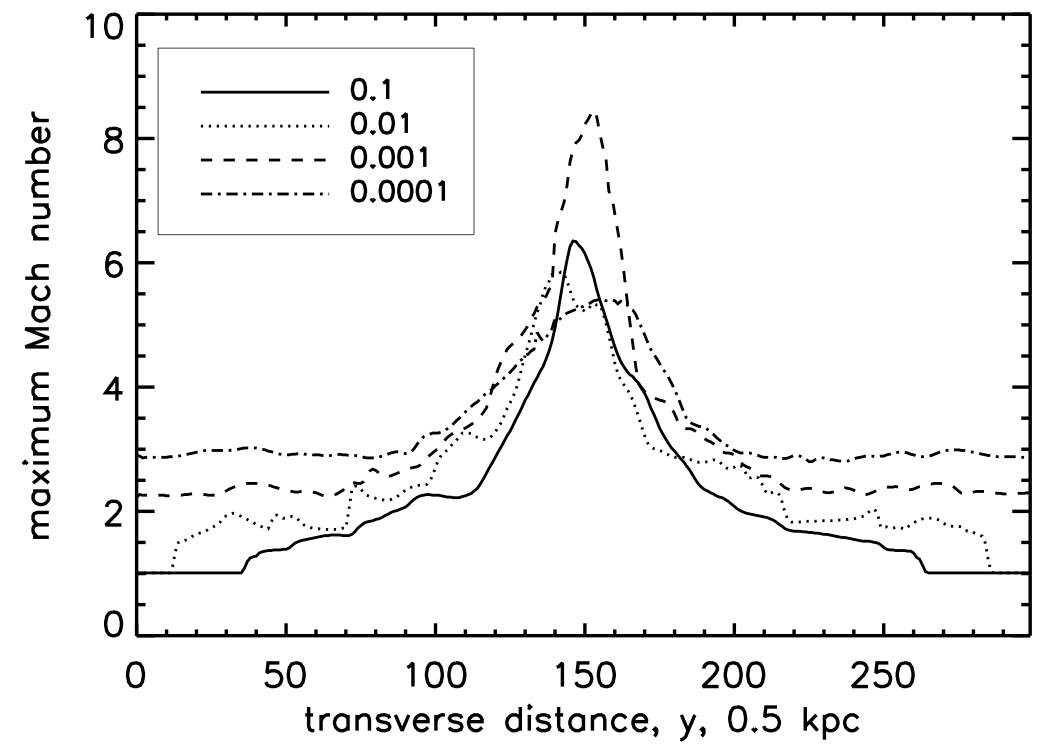}
\caption{ {\bf The Mach number of the bow shock in the cluster environment} as derived from the maximum jump in pressure taking steady shock jump conditions. The value is derived along each y-zone  in the z mid-plane. }
\label{machx}
\end{figure}

\subsection{Intracluster shock Mach number}

There are several useful Mach numbers that can be derived from the simulations. The first
 is from the physical properties. Here, we present the values for the time-averaged Mach numbers along the jet axis which  are easily derived for both the interface and the shock front running ahead of it.
These are listed in Columns 6 \& 7 of Table\,\ref{tab:model}. Note that these are generally high values but appropriate for estimating source ages. The bow cap is typically just 4--5\% ahead of the interface apart from the Mach 2 jet which has a 1\% stand-off and the anomalous precessing jet which appears as two distinct off-axis lobes encompassed by a surrounding large-scale bow. 

The Mach number attributed to the shock driven into the cluster medium can be derived from spectral measurements in the X-ray bands. One measure of the Mach number is calculated from the ratio of X-ray emission in the 1\, and 10\,keV bands. This generates an X-ray  temperature which can be compared to that expected from the jump conditions for a hydrodynamic shock.
These values, listed in the final column of Table\,,\ref{tab:model}, are very low since the extended low Mach number flanks of the bow dominate.

It can be concluded that the derived  Mach number depends on the resolution. This is demonstrated in Fig.\,\ref{machx} where shock hydrodynamic jump conditions have been employed to convert the maximum change in temperature or pressure along the x-axis into an equivalent Mach number for a steady shock. While a temperature jump may be measured observationally, in numerical simulation with an extreme density contrast, the pressure provides a better solution. This figure  demonstrates that any observation of the curved bow will be strongly dependent on the spatial resolution. Therefore, extreme caution must be applied when using derived Mach numbers to deduce radio galaxy energetics or dynamics.

\subsection{Evolution of X-ray flux density}

Can we identify candidate X-ray cavities from the total emission in the ambient medium? To determine this, we calculate the total Bremsstrahlung emission from the data cubes in the soft, medium and hard X-ray bands. Moreover,by presenting the time evolution, we gain a further potential diagnostic of radio galaxy age.
   
\begin{figure}
\includegraphics[width=0.47\textwidth]{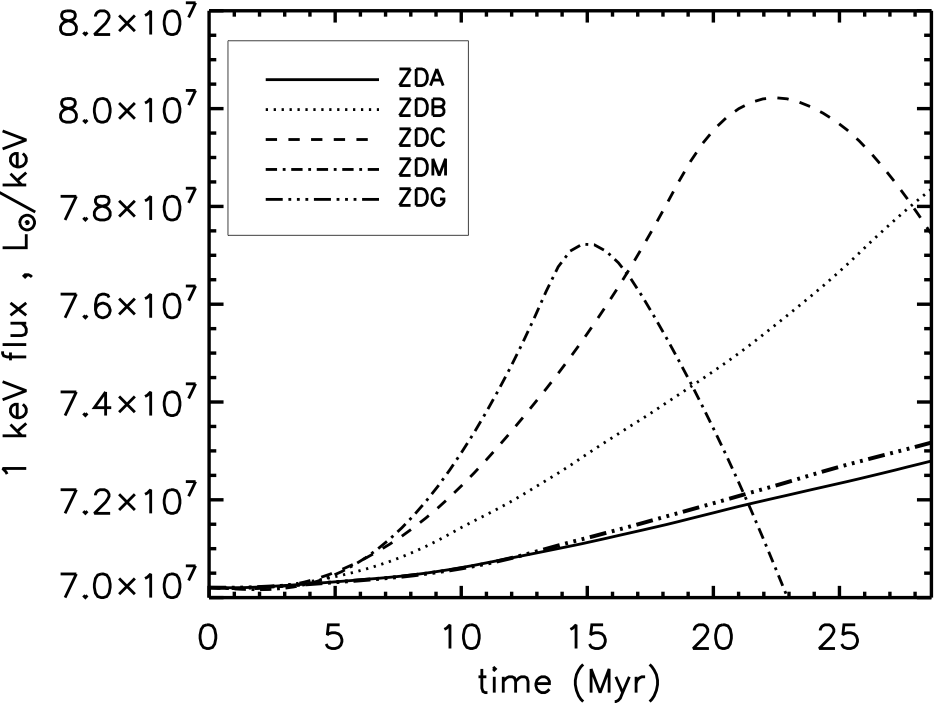}
\includegraphics[width=0.47\textwidth]{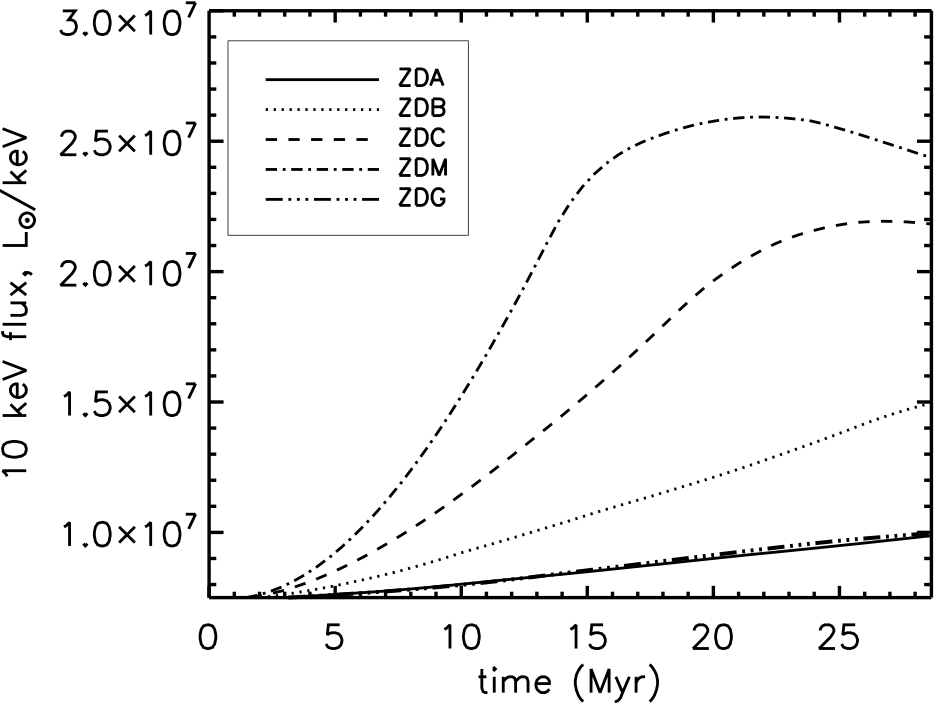}
\includegraphics[width=0.47\textwidth]{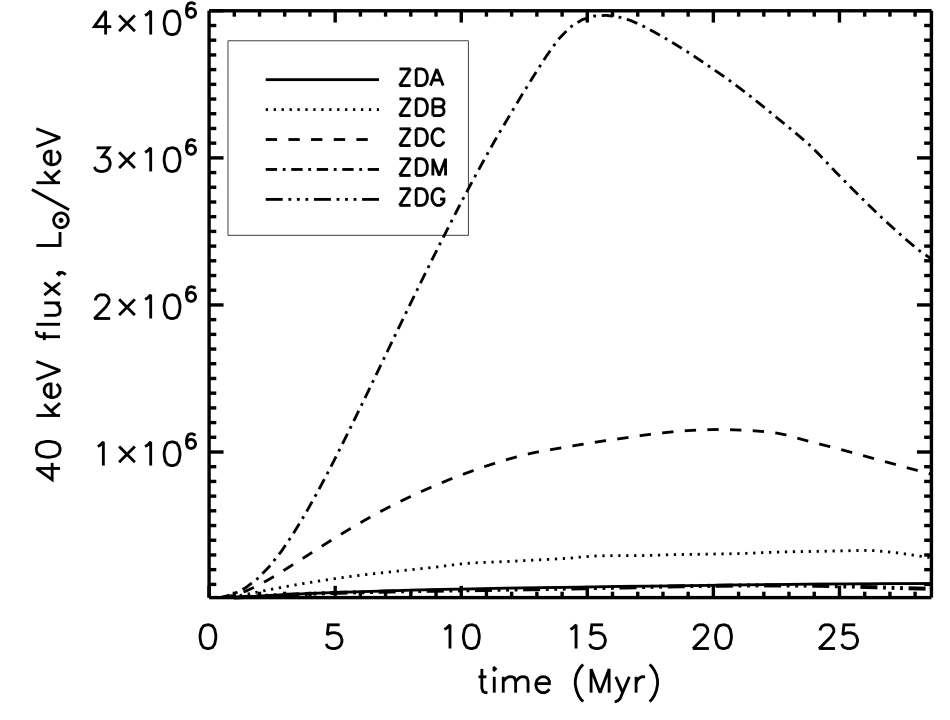}
\caption{ {\bf  Evolution of the flux density.} The total X-ray emission in each band of spectral width 1\,keV over the entire 150\,kpc cube as a function of time. }
\label{evolution}
\end{figure}

 Figure~\ref{evolution} shows that the total X-ray emission in each band increases despite the reduction within the cavity. This applies to all bands and is a result of the density squared dependence of the emissivity. Hence, when the radio lobes sweep up the material, they create significant density variations out of the ambient medium while the total mass remains unaltered. Therefore, the local depression is more than compensated by the enveloping region of enhanced emission.

This increase in total emission is mostly detectable in the harder bands. At 40\,keV, there is minimal emission before the jet is initiated and the emission strongly increases due to the direct impact. In this case, the emission is sensitive to the rise in temperature across the impact region. 

As the source ages, ambient material is ejected from the grid. This was found in the simulations with very low jet densities, as shown by the turn over in the emission curves. This is due to the large lateral expansion of the cavity. On the other hand, although a wide radio source is created in the case of a high jet precession angle, the 
X-ray emission is not significantly increased in comparison to the straight jet with $\eta = 0.1$.  We thus conclude that for the enhancement in integrated X-ray emission from the Bremsstrahlung process, the jet density is the crucial factor.

\section{Discussion}

A major  result is that the cavity morphology is a function of the jet-ambient density ratio. Note that this treats the jet Mach number and jet-ambient pressure ratio as the two other independent variables. We study here only the pressure-matched case and, for a fixed Mach number, this implies a constant momentum flow rate independent of how light the jet is. The pressure balance condition might be critical to the jet stability and will be relaxed in a follow-up study. Here, we find that for $\eta = 0.1$, an  ellipsoidal cavity demarcated by a shocked shell appears as an elliptical X-ray cavity and bright shell. As shown in Fig.\,\ref{ZDA-no-028-00}. the shell is typically 20\%--25\% thick and is likely to be asymmetrical, ultimately caused by a preferred direction of deflection of the jet in the leading impact region.

These cavity  results differ from those of   \citet{2015ApJ...803...48G}. A comparison is difficult since their simulations are two dimensional, involve viscosity and consider subsonic jets that generate top-wide cavities. The main difference, however, is that their  simulations only
considered cavities well after the jet had been switched off, allowing a buoyant bubble to rise in a galactic potential. 

As $\eta$ is reduced in the simulation sequence, the shell shape becomes progressively more conical with the high surface brightness regions 
located further into the flanks. Most remarkable is the appearance of ribs which cut across the cavity when the jet is in the plane of the sky. 
The ribs show up as distinct brightenings along the shell and finally as ripples at viewing angles close to the line of sight (fig.\,\ref{ZDB-yes-020-75}). This indicates that 
the structure is caused by oscillatory motion in the impact zone which has been known to experience periodic vortex shedding associated with 
periodic  transitions between double hotspots and single hotspots on the axis \citep{1985MNRAS.214...67S}. The shed vortex feeds back on 
to the approaching jet, squeezing it before the gas is shocked at the hotspot. The shedding time scale is 
estimated to be the hotspot radius  divided by the ambient sound speed. This yields four complete cycles for the $\eta = 0.01$ simulation, 
which is consistent with the simulated structure. This is confirmed through closer inspection of the radio image in the lower panel of 
Fig.\,\ref{ZDB-yes-023-00} which displays previously shed vortices on both sides of the radio cocoon, corresponding to the location of the X-ray ribs.

Enhanced X-ray structure within the Cygnus A thermal cocoon has been identified by \citet{2002ApJ...565..195S} as belts.  With {\em Chandra} data, the  description
was refined to cross-cutting thick ribs which appear to  be wrapped on the inside of an ellipsoidal shell \citep{2018MNRAS.476.4848D}. It is certainly reasonable to associate the observed ribs with the rib cage which occurs naturally in these simulations with the minimum of necessary ingredients. Further examples would be expected as 
we gain high resolution data for other classical doubles. 

For the lightest jets, the shell develops wide shoulders on the flanks and a deep  ovoid cavity. The radio jet penetrates through the cavity and propagates  across the domain.
These X-ray attributes are very similar to those associated with 3C\,444   \citep{2017MNRAS.466.2054V} and   3C\,320   \citep{2019MNRAS.485.1981V} which possess double  inner cavities with an adjacent rim plus outer shocks. 

Radio galaxies generated by jets within a wide precession cone would extend the lateral reach  of feedback into the intracluster medium. This is evident from the low Mach numer of the shocks into the cluster environment for the 20$^\circ$ precession angle cases in Table\,\ref{tab:model}. However, the fraction of jet energy transferred into the surroundings remains at a substantial level.
Therefore, precession does not actually  help to increase external support but 
may help spread this support of the gas across an entire cooling flow \citep{2019MNRAS.485.5590R} 
and so inhibit galaxy growth, two major issues recently summarised by \citet{2019A&A...622A..12H}. Precession may prove to be a prominent dynamical  mechanism in distant radio galaxies as they form in proto-clusters. Therefore, we concentrate below on the effects of precession  and the influence this may have on their apparent morphological classification.

 The dynamical times of the simulated radio galaxies are significantly shorter than the lifetime of a hot dense cooling cluster flow. 
This would indicate that the present active phase of an AGN would not 
be directly responsible for the onset of measurable feedback and increase of entropy in the cluster gas
 \citep{2007MNRAS.381.1109B}. Hence, it may be crucial to study how an environment responds after a radio galaxy power supply has been switched off.
  Indeed, cavities without enveloping shock waves are observed which suggest multiple episodes \citep{2004ApJ...606..185C} with shock decay in between.
   
An excellent template for an observed X-ray cavity and shell belongs to the environment of the fat double FR\.I 3C\,310 \citep{2012ApJ...749...19K}. 
    it is  a prime candidate for a radio galaxy formed by a fast precessing jet which has been able to draw out the radio hotspot into a ring in the southern lobe and a broken round arc in the northern lobe.
   In this case, the energetics have been clearly outlined  by \citet{2012ApJ...749...19K}.  The X-ray cavity and radio galaxy are powered by an ongoing outburst releasing
   thermal energy into the gas within the shocked inner 180\,kpc of 10$^{61}$
   ergs and which emits $\sim$ 2\,$\times$\,10$^{43}$\,erg\,s$^{-1}$  while the 
   outburst energy           is  $\sim$\,10$^{61}$\,ergs. The Mach number of 1.9 for the expansion is derived from a model for the shock density jump.  While these numbers are highly uncertain, this leads to an estimated time-averaged outburst  power of  2\,$\times$\,10$^{45}$\,erg\,s$^{-1}$ and suggests that the AGN could easily support the cooling flow with a duty cycle of just 10\%.

\begin{figure}
\includegraphics[width=0.5\textwidth]{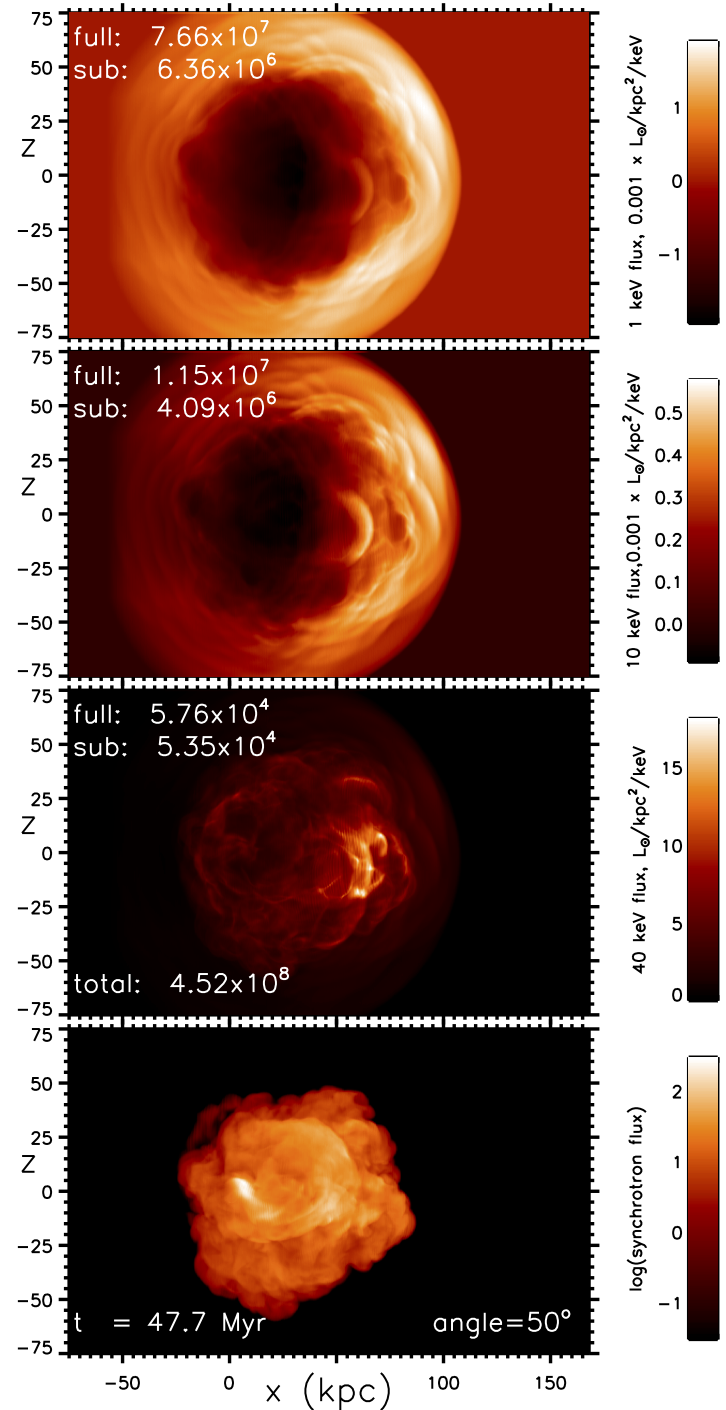}
\caption{ {\bf A fast precessing Mach 6  jet with 20$^\circ$ precession angle (for Model ZDG4).} This image is from the projected X-Z plane.  The jet-ambient density ratio is 0.1. The top three panels display the free-free X-ray surface brightness at 1\,keV, 10\,keV and 40\,keV   while the lower panel displays a radio image based on a simple synchrotron model. The jet axis is at an angle of
  50$^\circ$  to  the plane of the sky.}
\label{ZDG4-yes-047-50z}
\end{figure}

Here we employ our  model for the radio emission  of a fast precessing jet. The jet has a 
20$^\circ$ precession angle and a period of one quarter of the above standard models 
  (i.e. a period of $r_j/c_s$ where $r_j$ is the jet radius and $c_s$ is the ambient 
  sound speed. Figure\,\ref{ZDG4-yes-047-50z} displays a fat radio lobe with a ring-like 
  structure in the lobe apparent for this angle to the line of sight.  The total 
  X-ray power of the simulation of     2\,$\times$\,10$^{42}$\,erg\,s$^{-1}$  and jet 
  power of 6.4\,$\times$\,10$^{44}$\,erg\,s$^{-1}$ for this one half of the radio galaxy 
  within the 150\,kpc cube compares favourably without any need for parameter manipulation. 
  Note the outburst age of 48\,Myr and a derived time-averaged Mach number of the leading 
  edge of  1.44 for the interface and  2.37 for the disturbed ambient medium. These values 
  are commensurate with those derived from the jump conditions which yield a maximum 
  Mach number of 2.15. Note that this simulation was performed with extensions in the form of 
  a  staggered grid beyond the uniform displayed portion.

  X-ray cavities can take many forms and provide a means of investigating galaxy and cluster formation at high redshift since cluster cooling flows are strong X-ray sources. MACS J1447.4+0827 (z = 0.3755) provides a recent example of jet evacuated cavities  \citep{2020AJ....160..103P}. Unlike   Cygnus~A, in this case  two separate cavities on a scale of  100\,kpc are detected. Moreover, the cavities appear to open out away from the central galaxy. While this may be simulated here, the strong emission from the soft 
X-ray band occurs on the inner section of each side of the source. Thus we do not reproduce the distinct double cavity. This may well be remedied in simulations with, for example, ellipsoidal density profiles. On the other hand,
the double X-ray cavity associated with the  Phoenix galaxy cluster studied by \citep{2020PASJ...72...62A} consists of two well defined depressions that suggest an
interpretation in terms of  a low Mach number jet such as shown in 
Fig\,\ref{M2-yes-119-25}.

A number of cavities at low redshift do possess a double bubble structure with the LoFAR telescope revealing the driving radio galaxy \citep{2020MNRAS.496.2613B}. 
Examples are A1361 and MS0375). The single ovoid or thimble structures predicted here from the high Mach number jets are also found in this survey (e.g.   A2052, Zw8276) although less definitive. There is thus the indication that double bubbles are associated with samples derived from cluster surveys while ovals will be found ina association with powerful radio galaxies. 

  A cluster evolution scenario in which AGN feedback regulates the intracluster medium and so controls the development of central galaxies
  has  received much favourable attention \citep{2012ARA&A..50..455F,2019A&A...622A..12H}. 
    Here, we find quantitative support for the radio-mode in which 
  jet-induced outbursts provide repeated heating events in the environment
  \citep{2020MNRAS.tmp.1089G}. 
  Although we find that the jet power is effectively transferred into the cluster gas, one 10--50\,Myr outburst
  is only sufficient to provide a simmering of the immediately surrounding gas. Nevertheless, this is adequate to raise the energy in the gas 
  which can be transported through weak shocks and magnetohydrodynamic  waves into the entire cluster medium 
  on a ten times longer time scale in between the outbursts. Repeated bursts then act to raise the  energy until the 
  simmering gas is able to effectively cool and a cooling inflow develops. Without the repeated gentle simmering, the radio 
  burst would blow away the cluster medium. This implies that the cluster cooling flows we now see as strong X-ray sources have reached the point of balance between the   simmering process and the radiative cooling.    
   
\section{Conclusions}  
\label{conclusions} 
We have presented simulated X-ray images  of radio galaxies formed through supersonic hydrodynamic jets. Only synchrotron and thermal Bremsstrahlung processes have been considered without consideration for radiative losses or particle acceleration. Three dimensional simulations are necessary not only for the precession but also for the correct form of dissipation of turbulent energy through cascade downwards in wavelength.

Standard Mach 6 jets drive shells and cavities of roughly ellipsoidal shape and show a decrement in surface brightness of at most a few tens of per cent.
We have found that both  cavities/shells and bows/arcs are generated in the soft and hard X-ray bands, respectively. 

Straight FR\,II jets are associated with elliptical shells when the jet-ambient density ratio is high and with 
(projected as triangular) conical X-ray lobes when the jets are very light. Remarkable thick rib 
structures are evident when the jet-ambient density ratio is near 0.01.   
 
 Precession generates  bi-lobed or X-shaped radio sources and in curved or deflected radio jets.  In these cases, the strongest X-ray emission is found to be not associated with the hotspot but with the relic lobe or deflection location. This is because the regions of hot high-pressure gas and regions of dense high-compression  do not coincide. As the viewing angle is changed towards the line of sight, the cavity becomes a deep round  hole surrounded by circular ripples.   
  
Low Mach number shocks extend into the cluster environment but a single episode of radio galaxy formation is not sufficient for long-term support of a cooling flow. We have found that the total radio luminosity of a simulated radio galaxy does not increase with source age or size. This is because the integrated  emission is dominated by the higher pressure regions which are of relatively constant volume, rather than the low pressure lobes.
The jet kinetic power is mainly transferred into the ambient medium via shocks and compression as shown in Paper 1.  In contrast, the X-ray luminosity steadily increases, especially in the higher energy bands, until ambient gas is expelled from the simulated region.

A new result is the appearance of a rib cage in soft X-rays  which may be related to the structure found in Cygnus\,A as presented in detail from Chandra observations \citep{2018MNRAS.476.4848D}. We have only considered thermal X-ray emission in this work and so have investigated the shape of cavities.  It should also be noted that inverse Compton losses to photons from the Cosmic Microwave Background may well dominate the X-ray image as well as alter the radio structure and the morphological type as the relative contributions of the hotspot and diffuse lobe may systematically change \citep{2012MNRAS.420.2644K}. Furthermore, a non-thermal X-ray jet in Cygnus A
is mis-aligned from the radio jet suggesting a precession scenario \citep{2008MNRAS.388.1465S}. 
{
Relativistic jet flows are generally expected to yield similar flow patterns but light travel time effects may introduce new details. However, X-ray properties should not be altered. 
As noted above, the relativistic electrons which generate the radio emission are a source of energy which is transferred to cosmic microwave background photons via inverse Compton scattering throughout the lobes. This can fill in the cavities and reduce the number of radio galaxies which display cavities. 

Finally, using 3C\,310 as an example, we demonstrate that a radio galaxy is efficient at simmering the intracluster medium. A periodic gentle heating over several short episodes
could support the external medium through the propagation of weak shocks.

These conclusions are founded on several model assumptions.  
We assume a uniform ambient medium which will evidently never be the case and rarely valid as an approximation. The density within isolated galaxies and clusters are observed to decrease with distance from the bulge or core for nearby examples \citep{2008A&A...487..431C}. Moreover, for the distant radio galaxies being considered here, the propensity of progressing galaxy and cluster mergers are found to significantly disturb the density profiles and X-ray structures \citep{2017A&A...598A..61B}.

Adopting a non-uniform radial profile would have a profound influence on precessing jets as has been found for the near straight jet variety \citep{1997MNRAS.284..981C}. The jet would progress from light to heavy and the corresponding dominant radio structure change from the artist's brush to a ballistic pattern in which the jet path is delineated. For the external X-ray emission, the cocoon would become broader ad the radio galaxy penetrates into less dense fields. This reduces the emission fron the bow caps and creates considerably more open X-ray cavity structures, as often observed  and simulated \citep{2005A&A...431...45K}. 

In addition, it should be reiterated that we also assume no active influence from a magnetic field in either medium, non-relativistic speeds, no shear or spray on exit from the jet nozzle and an initial thermal pressure balance on entry.  All these assumptions will be tested in future works.

\section{Acknowledgements}  
\label{acks}

We wish to thank SEPnet for supplying infrastructure funding.

\section{Data availability}  
\label{data}
Data available on request. The data underlying this article will be shared on reasonable request to the corresponding author.

\bibliography{jets}
 \label{lastpage}

\end{document}